\documentclass[aps,prb,reprint,nofootinbib,longbibliography,superscriptaddress]{revtex4-2}
\usepackage{graphicx,graphics,color,epsfig}
\usepackage{bm}
\usepackage{dcolumn}
\usepackage{amsmath}
\usepackage{amssymb}
\usepackage{graphicx}
\usepackage{epstopdf}
\usepackage[colorlinks=true,  
    linkcolor=blue,          
    citecolor=blue,          
    urlcolor=blue,           
    anchorcolor=blue]{hyperref}
\usepackage{multirow}  
\usepackage{lineno} 

\newcommand{\be}{\begin{eqnarray}}
\newcommand{\ee}{\end{eqnarray}}

\newcommand{\ba}{\begin{align}}
\newcommand{\ea}{\end{align}}

\makeatletter

\newcommand{\Rmnum}[1]{\expandafter\@slowromancap\romannumeral #1@}
\makeatother

\def\sgn{\mathop{\rm sgn}}

\begin{document}

\title{ Optimized adiabatic-impulse protocol preserving Kibble-Zurek scaling \\ with attenuated anti-Kibble-Zurek behavior}

\author{Han-Chuan Kou}
\email{kouhc@swpu.edu.cn}
\affiliation{School of Sciences, Southwest Petroleum University, 637001, Nanchong, China}

\author{Zhi-Han Zhang}
\affiliation{College of Physics, Sichuan University, 610064, Chengdu, China}

\author{Xin-Hui Wu}
\affiliation{School of Sciences, Southwest Petroleum University, 637001, Nanchong, China}

\author{Yan Zhou}
\affiliation{School of Sciences, Southwest Petroleum University, 637001, Nanchong, China}

\author{Gang Chen}
\affiliation{School of Sciences, Southwest Petroleum University, 637001, Nanchong, China}

\author{Peng Li}
\email{lipeng@scu.edu.cn}
\affiliation{College of Physics, Sichuan University, 610064, Chengdu, China}

\date{\today}

\begin{abstract}
 We propose an optimized adiabatic-impulse (OAI) protocol that substantially reduces the evolution time for crossing a quantum phase transition while preserving Kibble–Zurek (KZ) scaling. 
 Near criticality, the control parameter is ramped linearly across the critical point at a rate characterized by a quench time $\tau_Q$. Away from criticality, the evolution remains adiabatic and is tuned close to the threshold of adiabatic breakdown, as quantified by an adiabatic coefficient $\zeta$ that scales as $\tau_Q^\alpha$. As a consequence, the total evolution time exhibits a sublinear power-law dependence on $\tau_Q$, and the conventional linear quench is recovered in the limit $\alpha\rightarrow\infty$. We apply the OAI protocol to the transverse Ising chain and numerically determine the minimal $\zeta$ required for KZ scaling. We further investigate the nonequilibrium dynamics in the presence of a noisy field that can induce anti–Kibble-Zurek (AKZ) behavior.
 Within the OAI protocol, noise-induced defects is significantly attenuated due to the shorter evolution time. The optimal quench time at which the defect density is minimized obeys an altered universal power-law scaling with the noise strength. Finally, we generalize the OAI protocol to the nonlinear quenches and numerically demonstrate a marked reduction in noise-induced defects.
\end{abstract}

\maketitle


\section{Introduction}

Quantum phase transitions play a central role in the equilibrium physics and non-equilibrium dynamics \cite{Sachdev_2011}. When a system is driven across a phase transition, nonequilibrium critical dynamics can induce the formation of topological defects, with a density that is predicted by the well-known KZ mechanism. This theory was first proposed by Kibble in the context of cosmology\cite{Kibble_1976, Kibble_1980} and discussed by Zurek in the thermal phase transitions of condensed matter systems \cite{Zurek_1985, Zurek_1993, Zurek_1996}. The KZ mechanism has been confirmed in a wide range of theories and experiments \cite{Polkovnikov_2008, Fischer_2010, Fischer_2010NJP, Zurek_2013, Singer_2013, Guo_2014, Carolyn_2016, Zurek_2018, Lara_2018, Sachdev_2019, Weinberg_2020, LiFx_2025}. In recent decades, the KZ mechanism has been also generalized to quantum phase transitions \cite{Damski_2005, Zurek_2005, Polkovnikov_2005, Dziarmaga_2005, Dziarmaga_2010}, and has become an important dynamical tool for measuring critical property in quantum simulation platforms \cite{ Sachdev_2019}.

The KZ mechanism can be attributed to the critical slowing down near the critical point, where the divergence of the relaxation time induces the appearance of topological defects regardless of how slowly the system is driven. The adiabatic-impulse approximation (AIA) framework captures this essential feature and provides a qualitative account of KZ scaling by comparing the relaxation time with the timescale induced by the external driving \cite{Zurek_2006, Fischer_2007, Polkovnikov_2011, Zurek_2016, Zurek_2017, GuoGC_2020, GaoXL_2021, Das_2021, Dziarmaga_2021, Zurek_2022, kou_2022, kou_2023, Dziramaga_2024, Yin_2025}. An extended AIA picture \cite{kou_2023} demonstrates that, as the driving rate increases, the defect density deviates from the KZ scaling and the dynamics enters pre-saturated and saturated regimes \cite{Dziarmaga_2013, Liu_2015, Shin_2019, Campo_2023, Campo_2023-3, Kiss_2025}. Reducing defects typically entails slower driving, thereby increasing the total evoulution time. However, a counterintuitive phenomenon, known as AKZ behavior, emerges due to the interaction with the environment and stochastic noise, leading to the production of more defects at slower driving rates \cite{Griffin_2012, Campo_2016, ZhuShiL_2017, Plenio_2020, Vicari_2020, Guo_2021, Suhas_2021, Campo_2022, Kastner_2023, Jara_2024, Ding_2024, Suzuki_2024, kou_2025, Jafari_2025}. 

In recent years, a variety of nonequilibrium protocols have been developed to speed up quenches across quantum phase transitions while avoiding additional defect production. Counterdiabatic driving has been proposed as a powerful tool that can, in principle, suppress nonequilibrium defects completely \cite{Campo_2012, Kazutaka_2013, Saberi_2014, Campbell_2022, Barone_2024, Suhas_2025}. This approach relies on an auxiliary Hamiltonian with nonlocal many-body interactions, which are difficult to implement in experimental platforms \cite{Damski_2014}. Several alternative routes are developed, including truncated counterdiabatic driving \cite{Campo_2012} and counterdiabatic optimized local driving approach \cite{Callum_2023, Barone_2024}.
For isolate system, nonlinear quench (NLQ) protocol can yield a lower defect density than the linear quench (LQ) protocol \cite{SenDiptiman_2008, Polkovnikov_2008NLQ, Dziarmaga_2024, Kastner_2026}. In finite-size systems, there exists an optimal NLQ protocol that minimizes the formation of defects \cite{WuN_2015}. By accounting for excited states, a nonadiabatic optimization scheme yields a lower defect density than the optimal NLQ protocol \cite{Campo_2025PRR}. 
In the quantum annealing regime, a faster annealing schedule was introduced to mitigate nonadiabatic excitations during target-state preparation \cite{Cerf_2002}, and a corresponding lower bound on the evolution time was derived in later developments \cite{Lucas_2023, Luis_2025}. The quantum adiabatic brachistochrone provides a time-optimal strategy \cite{Rezakhani_2009}, which can be determined more precisely by solving the corresponding brachistochrone equation \cite{WangXT_2015, DuJF_2016}.

Motivated by these investigations, we further analyze the AIA framework and design an OAI protocol that captures the KZ scaling. The construction is based on comparing the timescale $|\epsilon(t)/\dot{\epsilon}(t)|$ with the relaxation time $\tau(t)$, where $\epsilon(t)$ denotes a dimensionless distance from the critical point. 
As illustrated in Fig.~\ref{plot-AIA}, away from criticality the protocol chooses $\epsilon(t)$ such that $|\epsilon/\dot{\epsilon}|\propto\tau(t)$. This choice permits faster ramps further from the critical point while keeping the evolution adiabatic.
In the vicinity of the critical point, timescale $|\epsilon(t)/\dot{\epsilon}(t)|$ reduces to $|t|$. As a result, the OAI protocol substantially reduces the total evolution time while keeping the final defect density consistent with the KZ scaling. This further implies that, for dynamics governed by a Lindblad equation, the OAI protocol can yield a lower defect density than the LQ and NLQ protocols.

The remainder of this paper is organized as follows. In Sec. \ref{Sec-OD}, we formulate the OAI protocol based on the AIA framework. We analytically derive the scaling behavior of the total evolution time and the optimal quench time. The former is obtained as a function of quench time, whereas the latter is discussed in the presence of the noisy field. The OAI protocol is applied to the one-dimensional transverse Ising model in the absence and presence of a noisy field, as discussed in Sec. \ref{sec-Ising} and Sec. \ref{Sec-AKZ}, respectively. In Sec. \ref{Sec-NLOD}, the OAI protocol is generalized to incorporate the KZ scaling in nonlinear quenches. At last, a brief summary and discussion are given in Sec. \ref{Sec-Summary}.

\section{Optimized impulse-adiabatic protocol} \label{Sec-OD}

We start from the AIA framework, which is applicable to a variety of systems with second-order phase transitions \cite{Dziarmaga_2010}. We consider a time-dependent Hamiltonian
\begin{align}\label{Hgt}
  H(t)=H[g(t)] ,
\end{align}
where the control parameter is slowly ramped from $g_i \equiv g(t_i) $ to $g_f \equiv g(t_f)$ (with $g_i>g_f$) across the critical point $g_c$ with quench time $\tau_Q$. Here, $t_i$ and $t_f$ denote the initial and final time, respectively. A distance from a quantum critical point can be measured with a dimensionless parameter $\epsilon(t)=\frac{g(t)-g_c}{g_c}$. 

\begin{figure}[t]
  \begin{center}
		\includegraphics[width=3.3 in,angle=0]{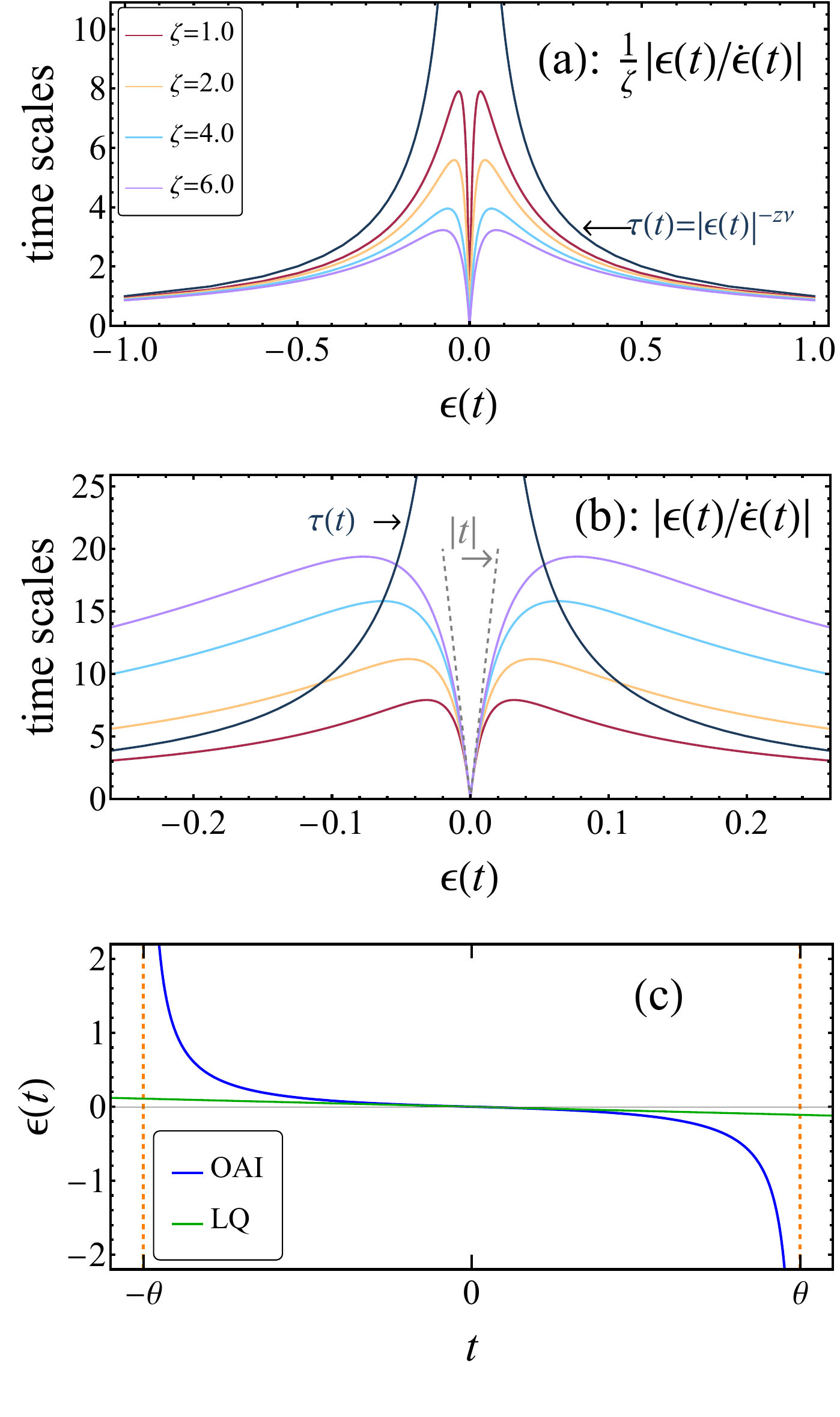}
  \end{center}
  \caption{ OAI protocol in the transverse Ising chain with fixed parameters $g_i=2$, $g_f=0$, and $\tau_Q=1000$. (a) and (b) show the two timescales, $|\epsilon(t)/\dot{\epsilon}(t)|$ and $\tau(t)$, for several representative values of $\zeta$, evaluated far from and near the critical point, respectively. 
  In (c), we compare the evolution of the dimensionless parameter $\epsilon(t)$ in the OAI protocol with that in the LQ protocol. In the OAI protocol, $\epsilon(t)$ diverges at $t=\pm\theta$, where the definition of $\theta$ is provided below Eq. (\ref{sol-epsi}). }
  \label{plot-AIA}
\end{figure}

In general, the system is initially prepared in the ground state of a simple Hamiltonian and subsequently driven to a final nontrivial state at which the topological defects induced by critical dynamics can be easily counted. For a conventional LQ protocol, 
\begin{align}
  \epsilon(t)=-\frac{t}{\tau_Q},
\end{align}
the AIA picture divides the dynamics into two adiabatic stages and an intermediate impulse stage by comparing the timescale $|\epsilon(t)/\dot{\epsilon}(t)|$ with the relaxation time $\tau(t)= |\epsilon(t)|^{-z\nu}$. Here, $z$ and $\nu$ are the dynamical and correlation length exponents, respectively.
In the adiabatic stages ($|\epsilon(t)|\gg0$), the system evolves adiabatically. Near the critical point, the critical slowing down causes the diverging relaxation time, so the system inevitably generates the topological defects or quantum excitations. The resulting defect density or excitation density is predicted by the KZ mechanism,
\begin{align}\label{SecI-KZexp}
  n\propto \left(\frac{J\tau_Q}{\hbar}\right)^{-\beta},
\end{align}
where $J$ sets the energy scale and $\beta=\frac{z\nu}{1+z\nu}$ is the KZ exponent. Throughout we set $\hbar=J=1$.

In this work, we propose the OAI protocol based on the AIA picture. In both adiabatic stages, we design $\epsilon(t)$ such that the driving timescale $|\epsilon(t)/\dot{\epsilon}(t)|$ is approximately proportional to the relaxation time $\tau(t)$
\begin{align}\label{secI-twotimes}
  |\epsilon(t)/\dot{\epsilon}(t)|\approx \zeta\tau(t)=\zeta|\epsilon(t)|^{-z\nu},~~|\epsilon(t)|\gg0,
\end{align}
as shown in Fig. \ref{plot-AIA}(a). 
We refer to $\zeta$ as the \emph{adiabatic coefficient} to mark the threshold of adiabatic breakdown.   
$\zeta$ is chosen to be positive and sufficiently large so that the evolution remains adiabatic in both adiabatic stages. In the two adiabatic stages, the time-dependent parameter $\epsilon(t)$ can be approximated as
\begin{align}\label{sol-epsilont1}
  \epsilon(t)\approx\epsilon_1(t)=\left\{
  \begin{array}{cc}
    \left[\frac{z\nu}{\zeta}(\theta+t)\right]^{-\frac{1}{z\nu}}, & t\ll 0, \\
    -\left[\frac{z\nu}{\zeta}(\theta-t)\right]^{-\frac{1}{z\nu}}, & t\gg 0 ,
  \end{array}\right.
\end{align}
where $\theta>0$ is a positive coefficient and the detail is presented in App. \ref{App-epsilont}. 

To remain consistent with the KZ dynamics where the final defect density satisfies Eq. (\ref{SecI-KZexp}), the system should be linearly driven across the critical point, 
\begin{align}\label{sol-epsilont2}
  \epsilon(t)\approx\epsilon_2(t)=-\frac{t}{\tau_Q}, &~~~ |\epsilon(t)|\sim 0,
\end{align}
Meanwhile, for $t\sim 0$, the part $(\theta\pm t)^{-x}$ in Eq. (\ref{sol-epsilont1}) can be expanded as $\theta^{-x}\pm x \theta^{-1-x}t+O(t^2)$. Combining Eqs. (\ref{sol-epsilont1}) and (\ref{sol-epsilont2}), we obtain the complete formulation of $\epsilon(t)$,
\begin{align}
  \epsilon(t)&=\epsilon_1(t)-\epsilon_1(0)\nonumber\\
  &=\left\{
  \begin{array}{cc}
    +\left(\frac{\zeta/z\nu}{\theta+t}\right)^{\frac{1}{z\nu}} -\left(\frac{\zeta}{z\nu \theta}\right)^{\frac{1}{z\nu}}, & t<0, \\
    -\left(\frac{\zeta/z\nu}{\theta-t}\right)^{\frac{1}{z\nu}} +\left(\frac{\zeta}{z\nu \theta}\right)^{\frac{1}{z\nu}}, & t\geq 0,
  \end{array}
  \right.
  \label{sol-epsi}
\end{align}
where $\theta=\frac{1}{z\nu}\left(\zeta^{\frac{1}{z\nu}}\tau_Q\right)^\frac{z\nu}{1+z\nu}$. A detailed derivation of Eq. (\ref{sol-epsi}) is provided in App. \ref{App-epsilont}. The term $\epsilon_1(0)$ should be small enough, 
\begin{align}\label{zeta-theta}
  \left.|\epsilon_1(0)\right.|=\left(\frac{\zeta}{z\nu \theta}\right)^{\frac{1}{z\nu}}=\left(\tau_Q^{-1} \zeta\right)^{\frac{1}{1+z\nu}} \ll 1,
\end{align} 
which ensures that the condition in Eq. (\ref{secI-twotimes}) remains satisfied.  
Accordingly, the OAI protocol requires an upper bound on $\zeta$, namely $\zeta\ll\tau_Q$. With an appropriate choice of $\zeta$, the adiabaticity in two adiabatic stages is maintained, and the whole nonequilibrium physics still is governed by the critical dynamics. The final defect density is given by KZ scaling, Eq. (\ref{SecI-KZexp}).

Equation~(\ref{sol-epsi}) indicates that the evolution is confined to the finite interval
\begin{align}
  -\theta\leq t_i<t_f\leq\theta.
\end{align}
Here, we set the adiabatic coefficient $\zeta$ as a power law of $\tau_Q$,  
\begin{align}\label{zeta-exponent}
  \zeta\propto \tau_Q^{\alpha}, ~~ \alpha <1.
\end{align} 
When Eq. (\ref{secI-twotimes}) holds, the above equation gives the upper bound of the exponent $\alpha$, while its lower bound is model dependent and must be determined for each system. As an example, we determine the lower bound of $\alpha$ for the transverse Ising model in Sec. \ref{Sec-II2}.
Then, the parameter $\theta=\frac{1}{z\nu} \tau_Q^{(\alpha+z\nu)/(1+z\nu)}$ is determined only by $\tau_Q$.
So, regardless of how distant the initial or final system is from the critical point, the total evolution time is bounded,
\begin{align}\label{TOD}
  T_\alpha \equiv t_f-t_i\leq 2\theta=\frac{2}{z\nu} \tau_Q^\frac{\alpha+z\nu}{1+z\nu}.
\end{align}
With this bound, the KZ scaling in Eq.~(\ref{SecI-KZexp}) can be expressed in terms of the total evolution time as $n_\text{KZ}\propto T_\alpha^{-d\nu/(\alpha+z\nu)}$.

When a noisy field is present, the Hamiltonian defined in Eq. (\ref{Hgt}) is modified to
\begin{align}\label{Hgamma}
  H_\gamma(t)=H[g(t)]+\gamma(t)V,
\end{align}
where $V$ is the noisy part. The resulting AKZ behavior induces more defects for slower driving \cite{Campo_2016}. Here, $\gamma(t)$ is Gaussian white noise with zero mean and the second moment $\langle\gamma(t)\gamma(t')\rangle=W^2\delta(t-t')$. Therefore, shortening the evolution time reduces the noise-induced contribution and hence yields lower defect density in the final state. When the OAI protocol is applied to Eq. (\ref{Hgamma}), the final defect density takes the form
\begin{align}
  n= a \tau_Q^{-\beta}+b \gamma T_\alpha
  =a \tau_Q^{-\beta}+b' \gamma \tau_Q^{\frac{\alpha+z\nu}{1+z\nu}},
\end{align}
where $a$, $b$, and $b'$ are three constants. There exists an optimal quench time 
\begin{align}\label{AKZ-tauQ}
  \tilde{\tau}_Q^{\{\text{OAI}\}}=\left(\frac{b'\beta \gamma}{a\alpha}\right)^{-\frac{1}{\alpha'+\beta}}\propto \gamma^{-\frac{1+z\nu}{(d+z)\nu+\alpha}},
\end{align}
which exhibits a universal scaling and corresponds to a minimum defect density $n$, where $\alpha'=\frac{\alpha+z\nu}{1+z\nu}$.

Finally, we emphasize three points.
First, the total evolution time defined in Eq. (\ref{TOD}) is rewritten as  $T_\alpha\propto\tau_Q$ for $\alpha>1$,  corresponding to $\zeta\gg\tau_Q$. In Eq. (\ref{zeta-exponent}), we have restricted the range of $\zeta$ to ensure the validity of Eq. (\ref{secI-twotimes}). The restriction, however, can be relaxed by allowing $\alpha>1$. Such a relaxation does not affect the adiabaticity of the adiabatic stages, but it renders Eq. (\ref{secI-twotimes}) invalid. Consequently, the resulting total evolution time is no longer the shortest time in the AIA framework.
Second, the LQ protocol is recovered as a limiting case of the OAI protocol. In the limit $\alpha\rightarrow\infty$, we have $\zeta, \theta\rightarrow\infty$ and Eq. (\ref{sol-epsi}) reduces to the LQ protocol by applying the approximation $(1-|t|/\theta)^{-1/z\nu}=1+|t|/(z\nu\theta)$. Detailed derivations supporting the first two points are presented in App \ref{app-ODandLQ}.
Third, we can rewrite Eq. (\ref{sol-epsi}) as 
\begin{align}
  \epsilon(t)
  &=\left\{
  \begin{array}{cc}
    +\left[\left(\frac{\zeta/z\nu}{\theta+t}\right)^{\frac{1}{z\nu}} -\left(\frac{\zeta}{z\nu \theta}\right)^{\frac{1}{z\nu}}\right]^r, & t<0, \\
    -\left[\left(\frac{\zeta/z\nu}{\theta-t}\right)^{\frac{1}{z\nu}} -\left(\frac{\zeta}{z\nu \theta}\right)^{\frac{1}{z\nu}}\right]^r, & t\leq 0,
  \end{array}
  \right.
  \label{sol-epsir}
\end{align}
In this case, the quench protocol can be reduced to the NLQ protocol near the critical point, $\lim_{t\rightarrow0}\epsilon(t)=-\sgn(t)|t/\tau_Q|^r$. The final defect density then takes the form
\begin{align}
  n\propto \tau_Q^{-\frac{rd\nu}{1+rd\nu}}.
\end{align}
We refer to Eq. (\ref{sol-epsir}) as the nonlinear OAI (NLOAI) protocol, whose numerical results in the concrete model are shown in Sec. \ref{Sec-NLOD}.

\section{Transverse Ising Chain under\\ the OAI Protocol}\label{sec-Ising}

As a prototypical model for a quantum phase transition, we consider the one-dimensional transverse Ising model,
\begin{align}\label{H-Ising}
H=-\sum_{j=1}^{N}\left( \sigma_{j}^{x}\sigma_{j+1}^{x}+g\sigma_{j}^{z}\right),
\end{align}
where $\sigma^{a}_j$ ($a=x,y,z$) are Pauli matrices and the total number of lattice sites $N$ is assumed to be even. We impose a periodic boundary condition, $\sigma^{a}_{N+j}=\sigma^{a}_{j}$, and focus on the ferromagnetic case. The strength of the transverse field is measured by $g$. Using the Jordan-Wigner mapping, $\sigma_j^z=1-2c_j^{\dagger}c_j$ and $\sigma_j^x=-(c_j^{\dagger}+c_j)\prod_{l<j}\sigma_l^z$, and the canonical Bogoliubov transformation, $c_{q}=u_{q}\eta_{q}-v_{q}\eta_{-q}^{\dagger}$ with the Bogoliubov coefficients $u_{q}$ and $v_{q}$, we can arrive at the diagonalized form of the Hamiltonian in the quasiparticle representation,
\begin{align}\label{even Hamiltonian}
H=\sum_{q}\omega_{q}(\eta_{q}^{\dagger}\eta_{q}-\frac{1}{2}),
\end{align}
where $\eta_{q}$ is the quasiparticle operator, $q$ the quasimomentum, and $\omega_{q}=2\sqrt{1+g^2-2g \cos q}$ the quasiparticle dispersion.

In the thermodynamic limit $N\rightarrow\infty$ and at zero temperature, there is a second-order quantum phase transition from a ferromagnetic state ($|g|<1$) with $\mathbb{Z}_{2}$ symmetry breaking to a quantum paramagnetic state ($|g|>1$) \cite{Sachdev_2011}. The QCP occurs at $g_c = 1$ or $g_c=-1$, where the quasiparticle dispersion becomes a linear one, $\omega_q\sim 2|q-q_c|$ with gap-closing mode $q_c=0$ or $\pi$, that is responsible for $z=\nu=1$.

\subsection{OAI protocol and KZ scaling}

Using the quench protocol in Eq. (\ref{sol-epsi}), we initialize the system in its ground state and ramp the transverse field from the paramagnetic ($g_i>1$) to the ferromagnetic phases ($0\leq g_f<1$) across a quantum critical point ($g_c=1$),
\begin{align}\label{equ-gt}
  g(t)=\left\{
  \begin{array}{cc}
    \frac{\zeta}{t+\sqrt{\zeta\tau_Q}}-\sqrt{\frac{\zeta}{\tau_Q}}+1, & t_i\leq t< 0, \\ [4pt]
    \frac{\zeta}{t-\sqrt{\zeta\tau_Q}}+\sqrt{\frac{\zeta}{\tau_Q}}+1, & 0\leq t\leq t_f, \\ 
  \end{array}
  \right.
\end{align}
where the initial and final times are given by $t_i=\frac{(1-g_i)\sqrt{\zeta}\tau_Q}{\sqrt{\zeta}-(1-g_i)\sqrt{\tau_Q}}$ and $t_f=\frac{(1-g_f)\sqrt{\zeta}\tau_Q}{\sqrt{\zeta}+(1-g_f)\sqrt{\tau_Q}}$, respectively. In Eq. (\ref{sol-epsi}), the coefficient $\theta$ is given by $\theta=\sqrt{\zeta\tau_Q}$. By adjusting the adiabatic coefficient $\zeta$, we maintain adiabaticity in the two adiabatic stages. Near the critical point, Eq. (\ref{equ-gt}) can be reduced to
\begin{align}\label{linearq}
  g(t)=1-\frac{t}{\tau_Q}+O(t^2),~~t\rightarrow 0,
\end{align}
where we adopt the approximation $\frac{1}{(t+C)^a}\approx \frac{1}{C^a}-\frac{at}{C^{1+a}}+O(t^2)$ for $t\rightarrow 0$. Fig. \ref{plot-AIA}(a)-(b) illustrate the evolution of two timescales, $|\epsilon/\dot{\epsilon}|$ and $\tau(t)$. Fig. \ref{plot-AIA}(c) depicts the evolution of the dimensionless parameter $\epsilon(t)$.

At the final time $t_f$, we set $g(t_f)=g_f=0$, so that the Hamiltonian in Eq. (\ref{H-Ising}) reaches the classical Ising limit and the system gets excited from its instantaneous ground state. The total number of defects (or kinks) can be measured by the operator,
\begin{equation}
  \mathcal{N} = \frac{1}{2}\sum_{j=1}^{N}\left(1-\sigma_j^x\sigma_{j+1}^x\right), \label{kinknumber}
\end{equation}
over the final state, which is in fact the number of excited quasiparticles \cite{Dziarmaga_2005}.
By mapping the dynamics to a series of Landau-Zener (LZ) problems \cite{Dziarmaga_2005}, the final defect density is related to the average excitation probability
\begin{equation}\label{define-n}
  n =\lim_{N\rightarrow\infty}\frac{1}{N} \langle\psi(t_f)|\mathcal{N}|\psi(t_f)\rangle=\frac{1}{\pi}\int_{q>0}dq~p_q.
\end{equation}
where $|\psi(t)\rangle$ is the quantum state at time $t$, and $p_q$ is the excitation probability. The detail is given by App. \ref{app-DTBdGE}.
For the slow quench, the standard KZ scaling takes the form
\begin{align}\label{n-KZS}
  n=n_\text{KZ}=\frac{1}{2\pi\sqrt{2\tau_Q}}.
\end{align}

\subsection{Numerical results and Adiabatic coefficient } \label{Sec-II2}

\begin{figure}[t]
  \begin{center}
		\includegraphics[width=3.3 in,angle=0]{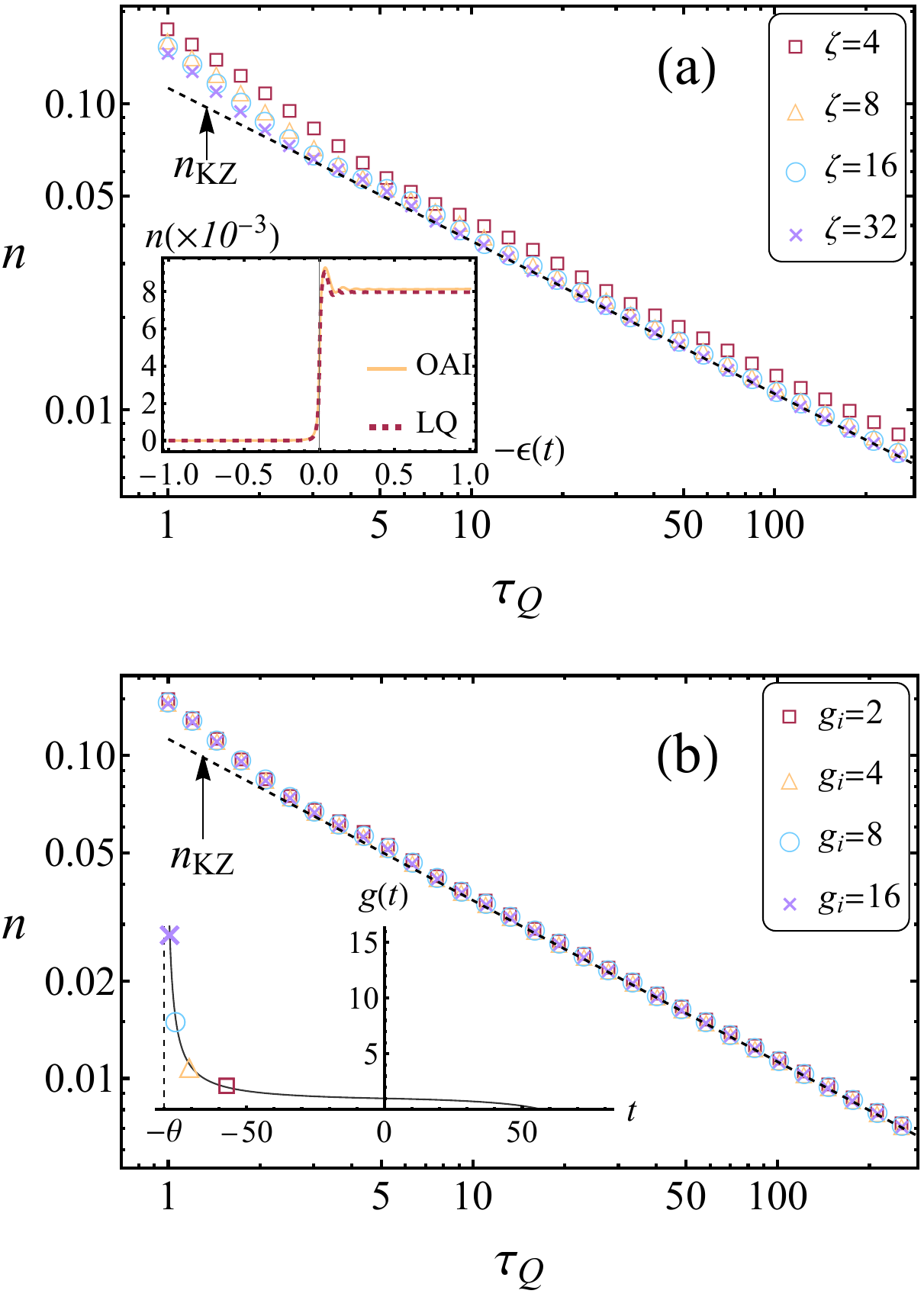}
  \end{center}
  \caption{Final defect density in the transverse Ising chain when the OAI protocol is applied. In (a), we set $g_i=2$ and $g_f=0$. The defect density (colored dingbats) gradually approaches the KZ scaling (black dashed line) as the adiabatic coefficient $\zeta$ increases. The inset of (a) compares the time evolution of the defect density for the OAI and LQ protocols at $\tau_Q=200$. For the OAI protocol we choose $\zeta=32$, while LQ protocol takes the form in Eq. (\ref{linearq}). The inset demonstrates the adiabaticity of the OAI protocol away from criticality. In (b), we fix $\tau_Q=200$, $\zeta=32$ and $g_f=0$, and the defect density data for different $g_i$ collapse onto the KZ scaling. The selected values of $g_i$ in (b) are indicated by the corresponding colored dingbats in the inset of (b).}
  \label{plot-defect}
\end{figure}

In Eq. (\ref{equ-gt}), we fix the adiabatic coefficient $\zeta$ and compute the final defect density by numerically solving the time-dependent Bogoliubov-de Gennes equations, Eq. (\ref{app-tdBdG}). 
In Fig. \ref{plot-defect}(a), we demonstrate that the final defect density (color dingbats) gradually falls into KZ scaling (black dashed line) as $\zeta$ increases (see $\zeta=4$, $8$, $16$ and $32)$. The inset of Fig. \ref{plot-defect}(a) depicts the time evolution of the defect density for the OAI and LQ protocols. It shows that, for a suitable $\zeta$, the OAI protocol maintains adiabaticity in both adiabatic stages, comparable to that of the LQ protocol.

Fig. \ref{plot-defect}(b) indicates, for a suitable $\zeta$, the final defect density follows the KZ scaling curve and is essentially independent of $g_i$. The inset depicts the evolution of $g(t)$ and marks several selected values of $g_i$ corresponding to the data sets in the main panel. We see that the OAI protocol accelerates the ramp away from criticality: the farther the system is from the critical point, the faster it is driven. 

However, a key question is what value of $\zeta$ is required to maintain adiabaticity in the two adiabatic stages. As shown in Fig. \ref{plot-defect}(a), the data show slightly deviation from the KZ scaling for $\zeta=4$, $8$, and $16$.
In the next part, we analyze how the adiabatic coefficient $\zeta$ affects the final defect density. 
First, in the limit $\zeta\tau_Q\rightarrow0$, the OAI protocol in Eq. (\ref{equ-gt}) can be viewed as a sudden quench where the final defect density is dominated by the initial condition \cite{kou_2023},
\begin{align}\label{n-sudden}
  n=n_\text{su}\approx \frac{1}{2}-\frac{1}{4g_i}, ~~\zeta\tau_Q\rightarrow0.
\end{align}

\begin{figure}[t]
  \begin{center}
		\includegraphics[width=2.85 in,angle=0]{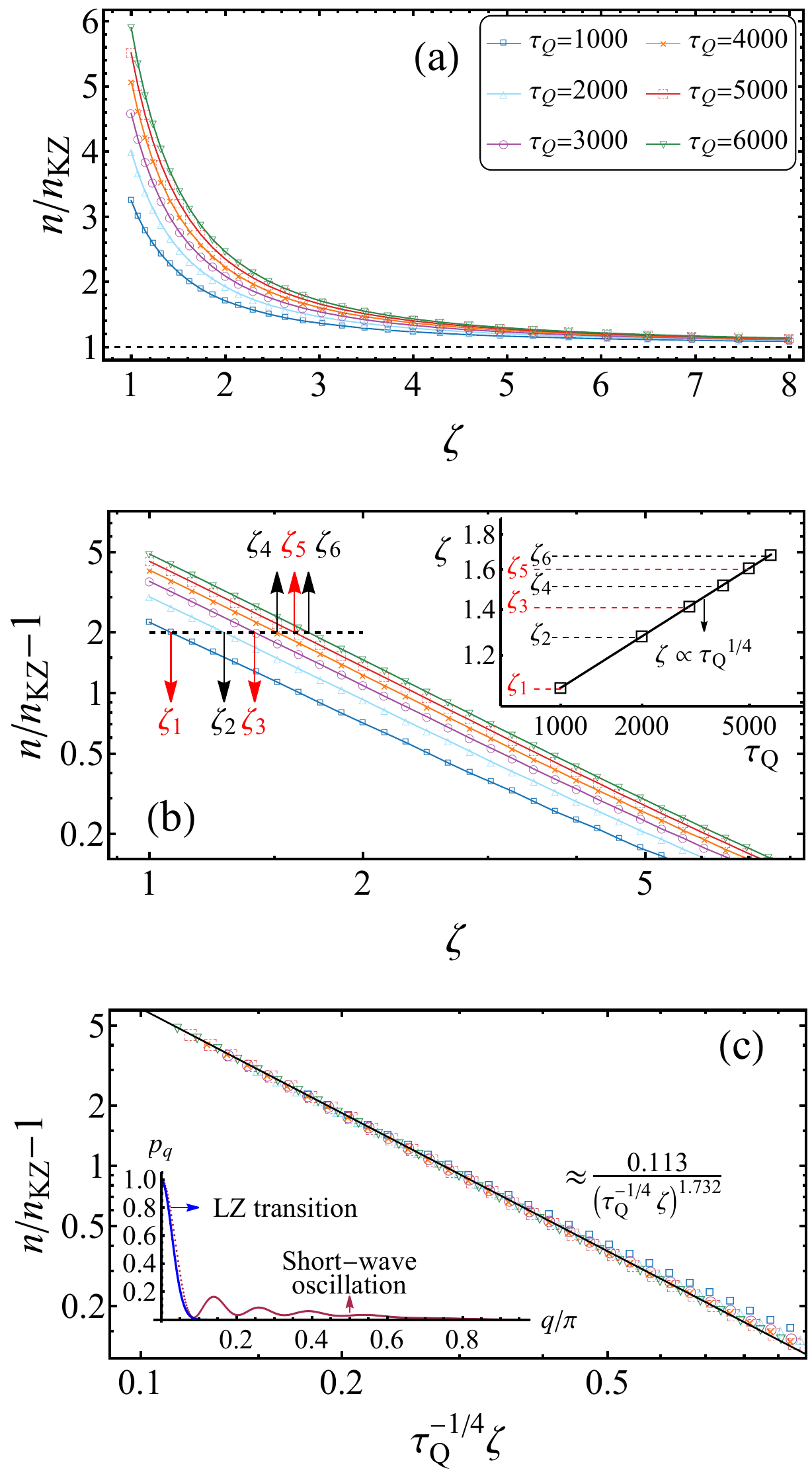}
  \end{center}
  \caption{ Scaled defect density as a function of $\zeta$ for several quench times, with $g_i=2$ and $g_f=0$. (a) demonstrates $n/n_\text{KZ}\rightarrow 1$ for sufficiently large $\zeta$. (b) presents a log-log plot of $n/n_\text{KZ}-1$ versus $\zeta$, revealing a power-law relationship.
  We fix $n/n_\text{KZ}-1=2$ (horizontal dashed line) and extract the corresponding intersections, yielding $\zeta_1=1.072$ ($\tau_Q=1000$), $\zeta_2=1.274$ ($\tau_Q=2000$), $\zeta_3=1.467$ ($\tau_Q=3000$), $\zeta_4=1.513$ ($\tau_Q=4000$), $\zeta_5=1.601$ ($\tau_Q=5000$), and $\zeta_6=1.675$ ($\tau_Q=6000$). The relationship between $\zeta$ and $\tau_Q$ is summarized in the inset of (b). By fitting these data, we approximately obtain $\zeta\propto\tau_Q ^{1/4}$. In (c), the horizontal axis is rescaled as $\tau_Q^{-1/4}\zeta$. The data reveal the scaling behavior of $F$ defined in Eq. (\ref{zeta-fit}), $F\approx 0.113(\tau_Q^{-1/4}\zeta)^{-1.732}$. In the inset of (c) depicts the excitation probability for $\tau_Q=10$ and $\zeta=0.8$, indicating that $F$ originates from short wave oscillations. }
  \label{plot-DDZeta}
\end{figure}

For finite $\zeta\tau_Q(\gg 0)$, Fig. \ref{plot-DDZeta}(a) shows that the scaled defect density $n/n_\text{KZ}$ depends on $\zeta$, where $n_\text{KZ}$ is defined in Eq. (\ref{n-KZS}). For $\zeta\gg 1$, the scaled defect density approaches $1$ (i.e., $n\rightarrow n_\text{KZ}$). Then, the log-log plot in Fig. \ref{plot-DDZeta}(b) demonstrates a power-law behavior between $\zeta$ and scaled defect density $n/n_\text{KZ}-1$,
\begin{align}\label{zeta-fit}
  n/n_\text{KZ}-1 = F^{-y}(\zeta,\tau_Q),
\end{align}
where the unknown function $F$ depends on $\zeta$ and $\tau_Q$. To determine $F$, we fix a reference value $n/n_\text{KZ}-1=2$, indicated by the horizontal dashed line in Fig. \ref{plot-DDZeta}(b). The intersections with the numerical curves then establish the corresponding pairs ($\zeta$, $\tau_Q$). The resulting $\zeta$ extracted as a function of $\tau_Q$ is shown in the inset of Fig.~\ref{plot-DDZeta}(b). A numerical fit indicates that $F$ scales as $F\propto\tau_Q^{-1/4}\zeta$. As shown in Fig. \ref{plot-DDZeta}(c), plotting $n/n_\text{KZ}-1$ versus the scaling variable $\tau_Q^{-1/4}\zeta$ yields a good data collapse for different values of $\tau_Q$. As a result, in the range of $0\ll \zeta\ll \tau_Q^{1/4}$, the defect density takes the form
\begin{align}\label{n-Zeta}
  n=n_\text{KZ}\left[ 1+x(\tau_Q^{-1/4}\zeta)^{-y} \right].
\end{align}
By numerical fitting, we have $x\approx 0.113$ and $y\approx 1.732$ for $g_i=2$ and $g_f=0$. Therefore, when $\zeta\geq \tau_Q^{1/4}$, the function $F$ in Eq. (\ref{zeta-fit}) becomes negligible, and the defect density follows the KZ scaling, $n=n_\text{KZ}$. Combining Eqs. (\ref{n-KZS}), (\ref{n-sudden}), and (\ref{n-Zeta}), we obtain the full expression of defect density:
\begin{align} \label{n-exp}
  n=\left\{
  \begin{array}{cc}
    n_\text{su}, & \zeta=0, \\
    n_\text{KZ}\left[ 1+x(\tau_Q^{-1/4}\zeta)^{-y} \right], & 0\ll \zeta\ll \tau_Q^{1/4},\\
    n_\text{KZ}, & \zeta\geq \tau_Q^{1/4}.
  \end{array}
  \right.
\end{align}

In the inset of Fig. \ref{plot-DDZeta}(c), we show the excitation probability $p_q$ as a function of $q$ for $\tau_Q=10$ and $\zeta=0.8$ (i.e., $\zeta\ll\tau_Q^{1/4}$). 
The term $n_\text{KZ}$ in Eq.~(\ref{n-Zeta}) is dominated by excitations of the gap-closing modes, for which the LZ transition, $p_{q\rightarrow 0}=e^{-2\pi\tau_Q q^2}$, provides an accurate description (blue curve). The remaining part in Eq. (\ref{n-Zeta}), proportional to $(\tau_Q^{-1/4}\zeta)^{-y}$, arises from the contribution of the short wave modes $0\ll q<\pi$, where $p_q$ exhibits pronounced coherent oscillations. These oscillations for short wave modes are rapidly suppressed as $\zeta$ increases. 

In Eq. (\ref{zeta-exponent}), we define the adiabatic coefficient as a function of quench time, $\zeta\propto\tau_Q^{\alpha}$. If we need a precise KZ scaling, the exponent $\alpha$ must be further restricted
\begin{align}
  \zeta\propto \tau_Q^\alpha, ~~\frac{1}{4}\leq \alpha <1,
\end{align}  
which ensures $\zeta\tau_Q^{-1/4}\geq 1$ and $\zeta\tau_Q^{-1}\ll 1$.

On the other hand, when the time-dependent system is driven in the presence of stochastic noise as shown in Eq. (\ref{Hgamma}), AKZ behavior induces higher defect density for slower driving rate. The defect density is minimized at a finite optimal quench time $\tilde{\tau}_Q$. In this noisy field setting,  we may take $\zeta$ to be a constant $C$ (i.e., $\alpha=0$) and choose $C$ sufficiently large,
\begin{align}\label{Czeta}
  \zeta= C\gg \tilde{\tau}_Q^{1/4},~~\alpha=0.
\end{align}
In the next section, the OAI protocol is applied in the presence of the noisy field, and we will discuss how $\zeta$ influences the scaling behavior of the optimal quench time dependence on the noisy strength.

\section{OAI Protocol in a noisy field}\label{Sec-AKZ}

In this section, we consider the system in the presence of the noisy field. The total Hamiltonian reads
\begin{align}\label{noise-Ising}
  H_\gamma(t) &= H(t)+\gamma(t)V,
\end{align}
where $H(t)$, defined in Eq. (\ref{H-Ising}), describes the noise-free Hamiltonian, and $V=\sum_{j=1}^N \sigma_j^z$ is the noise-dependent part. $\gamma(t)$ is a Gaussian white noise with zero mean and the nonzero second moment $\langle\gamma(t)\gamma(t')\rangle=W^2\delta(t-t')$, where $W$ characterizes the strength of the noise. Using Novikov’s theorem \cite{Novikov_1965}, the dynamics can be described by the Lindblad-type master equation, which is given by \cite{Campo_2016}
\begin{align}
  \frac{d}{dt}\rho(t)=-i[H(t),\rho(t)]-\frac{W^2}{2}[V,[V,\rho(t)]],
\end{align}
where $\rho(t)$ is the density operator where the initial condition reads $\rho(t_i)=|\psi(t_i)\rangle\langle\psi(t_i)|$.

By a Fourier transform, the density operator can be mapped to momentum space $\rho(t)=\bigotimes_{q>0}\rho_q(t)$. In each Fourier subspace, the Lindblad equation reads
\begin{align}
  \frac{d}{dt}\rho_q(t)=&-i[\sigma_zh_q^z(t)+\sigma_xh_q^x(t),\rho_q(t)]\nonumber\\
  &-\frac{W^2}{2}[\sigma_z,[\sigma_z,\rho_q(t)]],
\end{align}
where $h_q^z(t)=2g(t)-2\cos q$, $h_q^x=2\sin q$, and $g(t)$ is defined in Eq. (\ref{equ-gt}).

In the presence of noise, the defect generation is governed by two competing mechanisms: (i) noise-induced excitations, which increase with the noise strength $W^2$ and the total evolution time $T_\alpha$, and (ii) non-adiabatic excitations, which are suppressed by raising the quench time scale $\tau_Q$. As a result, the final defect density can be written as \cite{Campo_2016, kou_2025}
\begin{align}
   n=n_\text{KZ} + bW^2T_\alpha,
\end{align}
where $b$ is a constant and the total evolution time,
\begin{align}\label{TodIsing}
  T_\alpha=t_i-t_f \propto \tau_Q^{\frac{1+\alpha}{2}},
\end{align}
follows a sublinear power-law dependence on the quench time. This noise-induced contribution, $W^2T_\alpha$, gives rise to the AKZ behavior. The defect density can be rewritten
\begin{align}\label{AKZ}
   n=n_\text{KZ} + b\left(W^{\frac{4}{1+\alpha}}\tau_Q\right)^{\frac{1+\alpha}{2}}.
\end{align}
The defect density exhibits a minimum at a finite optimal quench time,
\begin{align}\label{tauQScaling}
   \tilde{\tau}_{Q}=\tilde{\tau}_{Q}^{\{\text{OAI}\}}\propto W^{-s},
\end{align}
with $s=\frac{4}{\alpha+2}$. The exponent $\alpha$ is restricted to $1/4\leq \alpha<1$, which is numerically demonstrated in Sec. \ref{Sec-II2}.

\begin{figure}[t]
  \begin{center}
		\includegraphics[width=3.3 in,angle=0]{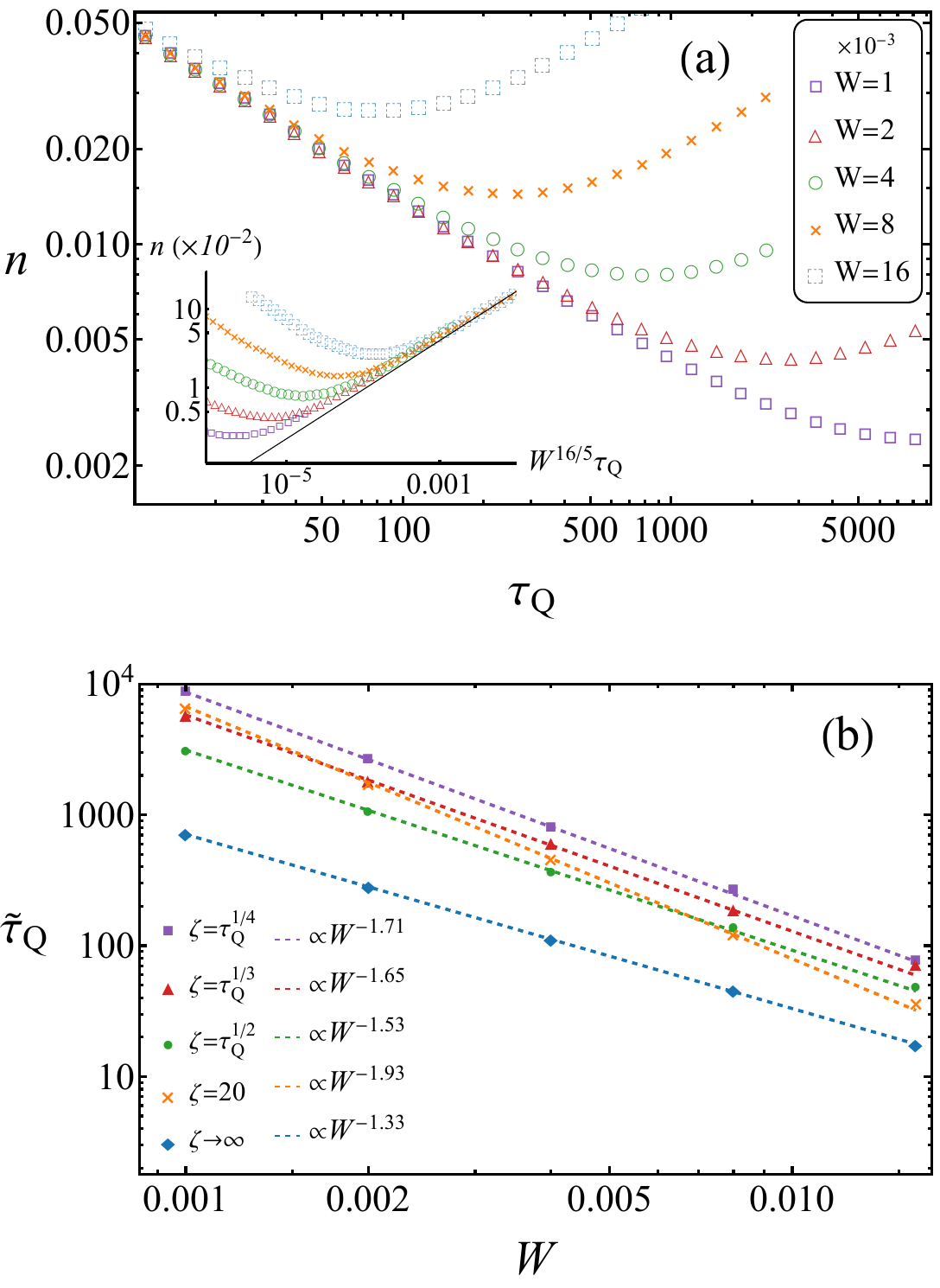}
  \end{center}
  \caption{ (a) Defect density $n$ versus $\tau_Q$ under a weak noisy control field, with $\zeta=\tau_Q^{1/4}$ (implying $\alpha=1/4$), $g_i=2$ and $g_f=0$. For sufficiently larger $\tau_Q$, the KZ scaling is suppressed and $n$ exhibits the AKZ behavior. In the inset of (a), the data collapse onto the black solid line given by Eq. (\ref{AKZ}), $n \propto (W^{16/5}\tau_Q)^{5/8}$. (b) depicts scaling behavior of the optimal quench time $\tilde{\tau}_Q$ versus $W$. Numerical results of $\tilde{\tau}_Q$ are shown as colored dingbats for different values of $\zeta$, while the corresponding fitting results are indicated by colored dashed lines. The data for $\zeta\rightarrow\infty$ are obtained by calculating Eq. (\ref{app-tdBdG}) in the LQ protocol. }
  \label{plot-AKZ}
\end{figure}

\begin{figure}[t]
  \begin{center}
		\includegraphics[width=3.28 in,angle=0]{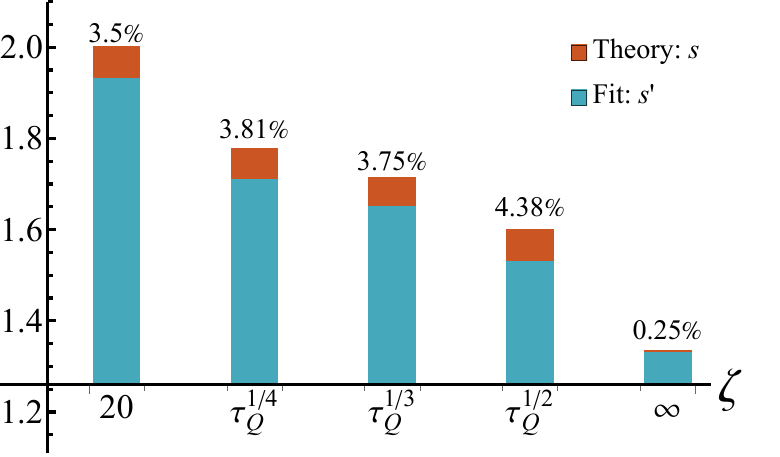}
  \end{center}
  \caption{ Bar chart comparing the scaling exponent of the optimal quench time $\tilde{\tau}_Q$ obtained from theory and numerical fitting. Burnt-orange bars show the theoretical prediction $s=\frac{4}{2+\alpha}$ defined in Eq. (\ref{tauQScaling}), while the corresponding exponent $s'$ is extracted from Fig. \ref{plot-AKZ}(b). The relative deviation $|s-s'|/s$ (in percent) is shown above each pair of bars. For $\zeta\ll\tau_Q$, the deviation is relatively small, approximately $4\%$. In the limit $\zeta\rightarrow\infty$, the OAI protocol reduces to the LQ protocol, and the deviation becomes negligible.}
  \label{plot-OptimaTau}
\end{figure}

In Fig. \ref{plot-AKZ}(a), we illustrate the defect density as a function of the quench time for $\zeta=\tau_Q^{1/4}$ and several selected $W$. For short quench times, the noise-induced contribution is negligible and the defect density follows the KZ scaling. With increasing quench time, the nonadiabatic dynamics is replaced by the AKZ behavior, where the defect density is governed by the noise-induced term in Eq. (\ref{AKZ}). As shown in the inset of Fig. \ref{plot-AKZ}(a), the data plotted versus the scaled quench time $W^{4/(1+\alpha)}\tau_Q$ are consistent with AKZ behavior.

The crossover from KZ scaling to AKZ behavior can be characterized by the optimal quench time, which we extract from the defect density $n$ at different noise strengths. In Fig. \ref{plot-AKZ}(b), we show the optimal quench time for several adiabatic coefficients. The data and corresponding fitted results represent by colored dingbats and colored dashed lines, respectively. It is observed that $\tilde{\tau}_Q$ decreases with increasing noise strength $W$, in good agreement with Eq. (\ref{tauQScaling}).
In practice, the optimal quench time is generally finite. So, we can fix the  adiabatic coefficient at a sufficiently large constant, e.g., $\zeta = 20$, as adopted in Eq. (\ref{Czeta}). In Fig. \ref{plot-AKZ}(b), the results for $\zeta=20$ are highlighted by the yellow dingbats.

As discussed at the end of Sec. \ref{Sec-OD}, the LQ protocol is fully recovered as the $\zeta \rightarrow \infty$ limit of the OAI protocol, for which the total evolution time reduces to $T = T_\infty \propto \tau_Q$. In this limit, the optimal quench time scales as
\begin{align}\label{tauQScalingLQ}
   \tilde{\tau}_{Q}=\tilde{\tau}_{Q}^{\{\text{LQ}\}}\propto W^{-\frac{4}{3}}.
\end{align}
For comparison, Fig. \ref{plot-AKZ}(b) also includes the numerical results for the LQ protocol ($\zeta\to\infty$, blue diamonds). In Fig. \ref{plot-AKZ}(b), the colored dashed lines show the fitting results of $\tilde{\tau}_Q$ for different values of $\zeta$. Notably, for the same noise strength $W$, the OAI protocol results in a significantly larger optimal quench time $\tilde{\tau}_Q$ than the LQ protocol.

The bar chart in Fig. \ref{plot-OptimaTau} compares the theoretical exponent $s=\frac{4}{2+\alpha}$ with the exponent $s'$ extracted from numerical fitting. The values of $s'$ are listed in the legend of Fig. \ref{plot-AKZ}(b).
We find a small deviation between the fitted exponent $s'$ and the theoretical prediction $s$ for the OAI protocol, while the deviation is negligible in the LQ limit ($\zeta\to\infty$). We quantify the relative deviation by $|s-s'|/s$ and annotate the corresponding percentages above each pair of bars. Obviously, the deviation is relatively small, approximately $4\%$ for several selected $\zeta$. 

\section{NLOAI protocol} \label{Sec-NLOD}

\begin{figure}[t]
  \begin{center}
		\includegraphics[width=3.3 in,angle=0]{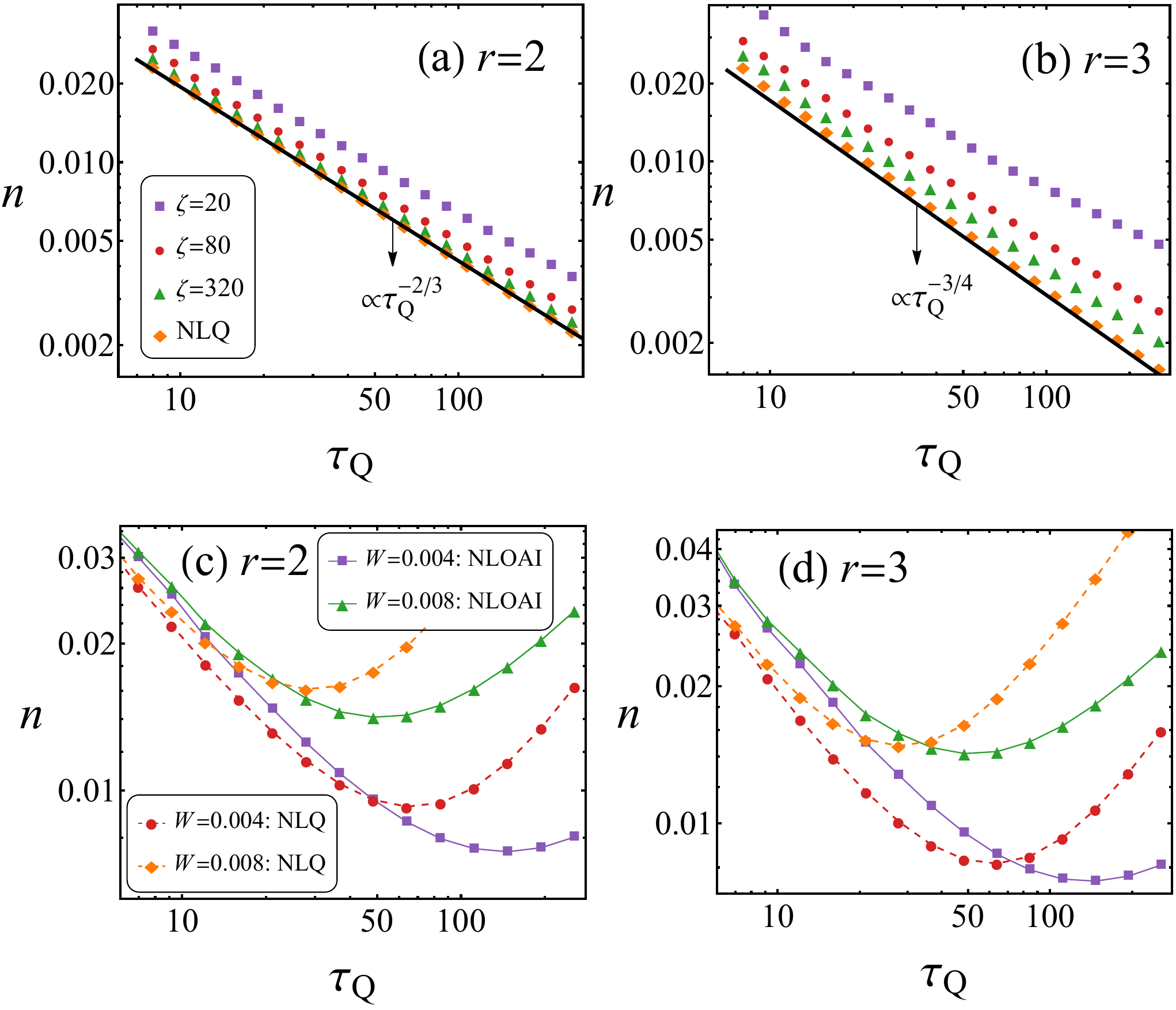}
  \end{center}
  \caption{ Final defect density in the transverse Ising chain when the NLOAI protocol is applied. We set $g_i=5$ and $g_f=0$. (a) shows noise-free numerical results for the defect density in the NLOAI and NLQ protocols for $r=2$, while (b) shows the corresponding results for $r=3$. The NLQ protocol is defined in Eq. (\ref{NLQ-protocol}). As $\zeta$ increases in the NLOAI protocol, the defect density gradually approach the theoretical prediction of Eq. (\ref{KZ-NLQ}). In (c) and (d), we illustrate the defect density for $r=2$ and $r=3$, respectively, with $\zeta=80$ at noise strength $W=0.004$ and $0.008$. Purple and green symbols denote the NLOAI results, while red and orange symbols denote the NLQ results. For large $\tau_Q$, the system applying NLOAI protocol generates a lower defect density than the system applying NLQ protocol.}
  \label{plot-NLOD}
\end{figure}

In Secs. \ref{Sec-OD} and \ref{Sec-AKZ}, we formulated an OAI protocol for the transverse field Ising chain, where the system is driven near the threshold of adiabatic breakdown in both adiabatic stages and crosses the critical point linearly. In this part, we generalize the OAI protocol to the NLOAI protocol,
\begin{align}\label{equ-NLOD}
  g(t)=\left\{
  \begin{array}{cc}
    +\left(\frac{\zeta}{\sqrt{\zeta\tau_Q}+t}-\sqrt{\frac{\zeta}{\tau_Q}}\right)^r+1, & t_i\leq t< 0, \\ [4pt]
    -\left(\frac{\zeta}{\sqrt{\zeta\tau_Q}-t}-\sqrt{\frac{\zeta}{\tau_Q}}\right)^r+1, & 0\leq t\leq t_f, \\ 
  \end{array}
  \right.
\end{align}
where the initial and final time read $t_i=\frac{-(g_i-1)^\frac{1}{r}\sqrt{\zeta}\tau_Q}{\sqrt{\zeta}+(g_i-1)^{\frac{1}{r}}\sqrt{\tau_Q}}$ and $t_f=\frac{(1-g_f)^{\frac{1}{r}}\sqrt{\zeta}\tau_Q}{\sqrt{\zeta}+(1-g_f)^\frac{1}{r}\sqrt{\tau_Q}}$, respectively. When the system crosses the critical point, Eq. (\ref{equ-NLOD}) can be reduced to a NLQ protocol
\begin{align}\label{NLQ-protocol}
  g(t)=-\sgn(t)\left|\frac{t}{\tau_Q}\right|^r+1, ~~t\to 0.
\end{align}
For sufficiently large adiabatic coefficient $\zeta$, the defect density in the NLOAI protocol takes the form
\begin{align}\label{KZ-NLQ}
  n\propto \tau_Q^{-\frac{dr\nu}{1+rz\nu}}.
\end{align}

In Fig. \ref{plot-NLOD}(a)-(b), we depict the defect density under the NLOAI and NLQ protocols for $r=2$ and $r=3$, respectively. The numerical results for the NLQ protocol (orange diamond) are in good agreement with Eq. (\ref{KZ-NLQ}), represented by the black curve. For the NLOAI protocol, when the adiabatic coefficient increases from $\zeta=20$, $80$ to $320$, the numerical results gradually approach the theoretical prediction of Eq. (\ref{KZ-NLQ}). In contrast to the OAI protocol, the NLOAI drives the system faster in both adiabatic stages, and thus requires a larger $\zeta$ to recover the KZ dynamics.

\begin{figure}[t]
  \begin{center}
		\includegraphics[width=3.28 in,angle=0]{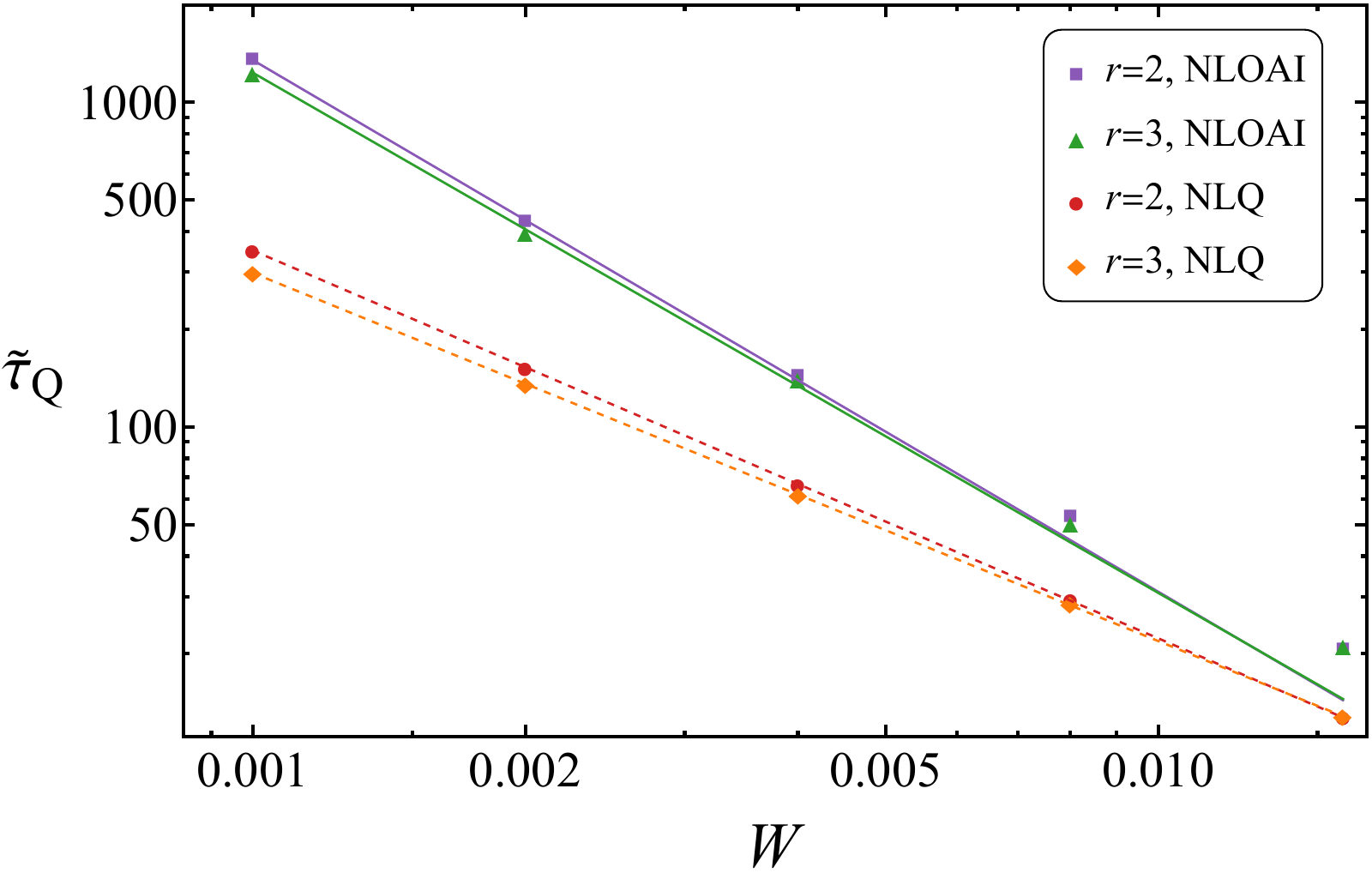}
  \end{center}
  \caption{ Scaling behavior of the optimal quench time $\tilde{\tau}_Q$. We fixed $g_i=5$, $g_f=0$ and $\zeta=80$. Numerical results for $\tilde{\tau}_Q$ are shown as colored dingbats for the NLOAI and NLQ protocols. Solid and dashed colored lines represent the corresponding fitting curves for the NLOAI and NLQ protocols, respectively. From top to bottom, the relative deviations $|s-s'|/s$ are $4.45\%$, $0.19\%$, $0.42\%$, and $0.78\%$, respectively.}
  \label{plot-NonL-TauQ}
\end{figure}

When a noise-free field is ramped from $g_i$ to $g_f$, the NLOAI protocol yields a slightly higher defect density than the NLQ protocol at finite $\zeta$, whereas it requires a shorter total evolution time. Consequently, in the presence of noise defined in Eq. (\ref{noise-Ising}), the dynamics under NLOAI protocol can in fact yield a lower defect density because it reduces the total exposure time to noise. As shown in Fig. \ref{plot-NLOD}(c)-(d), we fix $\zeta=80$ (i.e. $\alpha=0$) and consider $W=0.004$ and $W=0.008$. Obviously, the defect density in the NLOAI protocol (solid lines) is lower than that in the NLQ protocol (dashed lines). The optimal quench time scales as $\tilde{\tau}_Q\propto W^s$, with
\begin{align}
  s=\frac{4(1+r)}{1+3r}
\end{align}
as shown in Fig. \ref{plot-NonL-TauQ}. The relative deviation $|s-s'|/s$ between theoretical exponent $s$ and fitting exponent $s'$ is about $4\%$ for both quench protocol.

\section{Summary and Discussion}\label{Sec-Summary}

In summary, we propose an OAI protocol that realizes the much shorter evolution time within the AIA framework for systems whose dynamics is governed by the KZ mechanism. We define an adiabatic coefficient $\zeta$ that characterizes the threshold for adiabatic breakdown in the two adiabatic stages. Applying the OAI protocol to the transverse Ising chain in the absence and presence of stochastic noise reveals a new scaling of the optimal quench time. The resulting universal scaling law suggests that the OAI protocol can be generalized, making it a suitable tool for studying nonequilibrium critical dynamics in other integrable and nonintegrable systems.

In D-Wave quantum annealers, a non-adiabatic process during fast evolution prevents the preparation of the adiabatic quantum state \cite{ Hauke_2020, Crosson_2021, Fioroni_2025}. An appropriate evolution protocol can therefore provide an efficient means of satisfying the adiabatic condition during quantum annealing \cite{Zurek_2018, Munoz_2025}. 
For a finite system of linear size $L$, the adiabatic condition requires that $\tau_Q\sim \frac{1}{\varepsilon} L^\frac{1+z\nu}{\nu}$ in order to maintain the defect density less than $\varepsilon$ \cite{Dziarmaga_2010, Boris_2010}. 
Within the OAI protocol, the total evolution time required to satisfy this adiabatic condition follows directly from Eq. (\ref{TOD}),
\begin{align}
  T_\alpha(L)\sim \varepsilon^{-\frac{1+\alpha}{2}}L^{\frac{(1+z\nu)(1+\alpha)}{2\nu}}, ~~\alpha<1.
\end{align}
This result provides an analytically tractable route to accelerate the driving process while maintaining the adiabatic condition, and may thus offer useful guidance for more efficient adiabatic state preparation.

\section{ACKNOWLEDGMENTS}

This work is supported by the Natural Science Foundation of Sichuan Province (2026NSFSC0772) and the National Natural Science Foundation of China (11074177).

\appendix

\section{Formulation of the OAI protocol } \label{App-epsilont}

This Appendix provides the detail about the OAI protocol.
As shown in Fig. \ref{plot-AIA}(c), the control parameter is ramped across the critical point with $\dot{\epsilon}(t)<0$ and
\begin{align}
\left\{
\begin{array}{cc}
  \epsilon(t)>0~~\text{and} & t< 0, \\
  \epsilon(t)\leq 0~~\text{and} & t\geq 0 .
\end{array}
\right.
\end{align}
In Eq. (\ref{secI-twotimes}), we give the relation between the driving timescale $|\epsilon/\dot{\epsilon}|$ and relaxation time $\tau(t)$ for $|\epsilon|\gg0$. Here, we introduce an auxiliary variable $\epsilon_1(t)$ that meets 
\begin{align}\label{app1-deltaepsilon}
  \epsilon(t)=\epsilon_1(t)+\delta_\epsilon,
\end{align} 
and
\begin{align}\label{app1-epsi1t}
|\epsilon_1(t)/\dot{\epsilon}_1(t)|=\zeta\tau(t),
\end{align}
where $\zeta(>0)$ is the adiabatic coefficient and $\delta_\epsilon$ is a sufficiently small offset. Eq. (\ref{app1-epsi1t}) can be rewritten as
\begin{align}
  \left\{
  \begin{array}{cc}
    \epsilon_1^{1+z\nu}=-\zeta\dot{\epsilon_1},& t< 0,\\
    (-\epsilon_1)^{1+z\nu}=-\zeta\dot{\epsilon_1},& t\geq0.
  \end{array}
  \right.
\end{align}
Then, we can write the solution of $\epsilon_1(t)$,
\begin{align}\label{app1-epsi1}
  \epsilon_1(t)=-\Theta'(t)\left[\frac{z\nu}{\zeta}\left(\theta-|t|\right)\right]^{-\frac{1}{z\nu}},
\end{align}
where $\theta$ is a positive coefficient, and the condition $\theta-|t|>0$ restricts the evolution time to the range of
\begin{align}
   -\theta\leq t_i<t_f\leq\theta.
\end{align}
We have $\Theta'(t)=2\Theta(t)-1$ where $\Theta(t)$ is Heaviside step function, $\Theta(t<0)=0$ and $\Theta(t\geq 0)=1$.

Near the critical point $t\to 0$, $\epsilon_1(t)$ can be written as
\begin{align}\label{app1-solue1}
  \epsilon_1(t)=-\Theta'(t)\left(\frac{\zeta}{z\nu \theta}\right)^{\frac{1}{z\nu}}
  -\frac{\zeta^{\frac{1}{z\nu}}}{(z\nu \theta)^{1+\frac{1}{z\nu}} }t+O(t^2).
\end{align}
To ensure a linear crossing of the critical point, we set  $\delta_\epsilon=\epsilon_1(0)$. And then, we can obtain the complete formulation of $\epsilon(t)$, 
\begin{align}\label{app1-et}
  \epsilon(t)&=\epsilon_1(t)-\epsilon_1(0)=
  \epsilon_1(t)+\Theta'(t)\left(\frac{\zeta}{z\nu \theta}\right)^{\frac{1}{z\nu}}.
\end{align}
By comparing with Eq. (\ref{app1-solue1}) and the linear quench in Eq. (\ref{sol-epsilont2}), we have
\begin{align}\label{app1-tauQ}
  \frac{\zeta^{\frac{1}{z\nu}}}{(z\nu \theta)^{1+\frac{1}{z\nu}} }=\frac{1}{\tau_Q}.
\end{align}
This condition determines $\theta$ as
\begin{align}\label{app-theta}
  \theta=\frac{1}{z\nu}\left(\zeta^{\frac{1}{z\nu}}\tau_Q\right)^\frac{z\nu}{1+z\nu}.
\end{align}

\section{LQ protocol as a limit of the OAI protocol}\label{app-ODandLQ}

When the initial and final parameters are two finite values, $\epsilon_i\equiv\epsilon(t_i)(>0)$ and $\epsilon_f\equiv\epsilon(t_f)(<0)$, the initial and final times are given by
\begin{align}\label{app-ti1}
  t_i=-\theta+\theta\left[1+\epsilon_i\left(\frac{\zeta}{z\nu\theta}\right)^{-1/z\nu}\right]^{-z\nu},
\end{align}
and 
\begin{align}\label{app-tf1}
  t_f=\theta-\theta\left[1-\epsilon_f\left(\frac{\zeta}{z\nu\theta}\right)^{-1/z\nu}\right]^{-z\nu},
\end{align}
respectively.
Then, in Eq. (\ref{zeta-exponent}), we set the adiabatic coefficient to scale with quench time, $\zeta=\tau_Q^\alpha$. Therefore, the total evolution time is restricted by Eq. (\ref{TOD}), $T_\alpha=t_f-t_i\leq 2\theta$.

Using Eq. (\ref{zeta-exponent}), we obtain
\begin{align}\label{app-zeta-theta}
  \frac{\zeta}{\theta}=z\nu\tau_Q^{\frac{z\nu(\alpha-1)}{1+z\nu}},
\end{align} 
and
\begin{align}
  \left\{
  \begin{array}{cc}
    \zeta\ll\theta, & \alpha<1 ~(\text{i.e.,}~\zeta\ll\tau_Q), \\
    \zeta\gg\theta, & \alpha>1 ~(\text{i.e.,}~\zeta\gg\tau_Q).
  \end{array}
  \right.
\end{align}
In the case of $\alpha>1$, Eqs. (\ref{app-ti1}) and (\ref{app-tf1}) simplify to
\begin{align}\label{app2-ti}
  t_i\approx-\theta+\theta\left[1-\epsilon_iz\nu\left(\frac{\zeta}{z\nu\theta}\right)^{-1/z\nu}\right]=-\epsilon_i\tau_Q,
\end{align}
and 
\begin{align}\label{app2-tf}
  t_f\approx\theta-\theta\left[1+\epsilon_fz\nu\left(\frac{\zeta}{z\nu\theta}\right)^{-1/z\nu}\right]=-\epsilon_f\tau_Q,
\end{align}
respectively. Therefore, the total evolution time is turn into 
\begin{align}
  T_\alpha\approx(-\epsilon_f+\epsilon_i)\tau_Q.
\end{align}

Moreover, in the large-$\zeta$ limit ($\zeta\to\infty$), Eq.~(\ref{app1-et}) simplifies to
\begin{align}\label{app2-epsi}
   \lim_{\zeta\rightarrow\infty}\epsilon(t)=-\Theta'(t)\left(\frac{\zeta}{z\nu\theta}\right)^{\frac{1}{z\nu}}\frac{|t|}{z\nu\theta},
\end{align}
since $|t|\ll\theta$ for any finite $t$. In Eqs. (\ref{app2-ti}), (\ref{app2-tf}) and (\ref{app2-epsi}), we adopt the approximation, $(1-|t|/\theta)^{-1/z\nu}=1+|t|/(z\nu\theta)$. Utilizing Eq. (\ref{app1-tauQ}), the OAI protocol can be reduced to the LQ protocol,
\begin{align}
  \lim_{\zeta\rightarrow\infty}\epsilon(t)=-\Theta'(t)\frac{|t|}{\tau_Q}=-\frac{t}{\tau_Q}.
\end{align}

\section{Time-dependent Bogoliubov-de Gennes equations}\label{app-DTBdGE}

As time evolves, the quantum state $|\psi(t)\rangle$, which gets excited from the instantaneous ground state, should follow the time-dependent Bogoliubov transformation
\begin{align}
c_{q}=u_{q}(t)\tilde{\eta}_{q}+v^{*}_{-q}(t)\tilde{\eta}_{-q}^{\dagger},
\end{align}
where the quantum state has to be annihilated by the Bogoliubov fermions $\tilde{\eta}_q$ at every instant: $\tilde{\eta}_q|\psi(t)\rangle=0$. In Heisenberg picture, the fermion operator and Bogoliubov quasiparticle operator should satisfy  $i\frac{\mathrm{d}}{\mathrm{d}t}\tilde{\eta}_{q}=0$ and $i\frac{\mathrm{d}}{\mathrm{d}t}c_{q}=\left[c_{q}, H\right]$ \cite{Dziarmaga_2005}.

we can arrive at the dynamical version of the time-dependent Bogoliubov-de Gennes equations,
\begin{align}\label{app-tdBdG}
i\frac{\mathrm{d}}{\mathrm{d}t}
\left[
\begin{array}{ccc}
u_{q}(t)\\v_{q}(t)
\end{array}
\right] =
\left[
\begin{array}{ccc}
\epsilon_{q}(t)&\Delta_{q}\\\Delta_{q}&-\epsilon_{q}(t)
\end{array}
\right]
\left[
\begin{array}{ccc}
u_{q}(t)\\v_{q}(t)
\end{array}
\right],
\end{align}
where $\epsilon_q(t)=2\{g(t)-\cos q\}$ and $\Delta_q=2\sin q$. It can be solved exactly by mapping to the LZ problem \cite{Dziarmaga_2005}. When the OAI protocol is applied, we need to solve this problem and calculate the density of defects through the excitation probability in the final state of the system.

And then, the LZ excitation probability is given by
\begin{align}\label{define-pq}
  p_q(t) =& \langle\psi(t)|\eta^\dagger_q\eta_q|\psi(t)\rangle
\end{align}
at any time $t$.


%

\end{document}